# Towards quantum-limited coherent detection of terahertz waves in charge-neutral graphene

S. Lara-Avila[1,2*], A. Danilov[1], D. Golubev[3], H. He[1], K.H. Kim[1], R. Yakimova[4], F. Lombardi[1], T. Bauch[1], S. Cherednichenko[1], S. Kubatkin[1]

[1]Department of Microtechnology and Nanoscience, Chalmers University of Technology, SE-412 96, Gothenburg, Sweden

[2]National Physical Laboratory, Hampton Road, Teddington, TW11 0LW, UK

[3]Department of Applied Physics, Aalto University, FI-00076 Aalto, Finland

[4]Department of Physics, Chemistry and Biology, Linkoping University, S-581 83, Linköping, Sweden.

**Spectacular advances in heterodyne astronomy with both the Herschel Space Observatory[1] and Stratospheric Observatory for Far Infrared Astronomy (SOFIA)[2] have been largely due to breakthroughs in detector technology[3]. In order to exploit the full capacity of future THz telescope space missions (e.g. Origins Space Telescope[4]), new concepts of THz coherent receivers are needed, providing larger bandwidths and imaging capabilities with multi-pixel focal plane heterodyne arrays[5]. Here we show that graphene, uniformly doped to the Dirac point, enables highly sensitive and wideband coherent detection of THz signals. With material resistance dominated by quantum localization, and thermal relaxation governed by electron diffusion, proof-of-concept graphene bolometers demonstrate a gain bandwidth of 8 GHz and a mixer noise temperature of 475 K, limited by residual thermal background in our setup. An optimized device will result in a mixer noise temperature as low as 36 K, with the gain bandwidth exceeding 20 GHz, and a Local Oscillator power of < 100 pW. In conjunction with the emerging quantum-limited amplifiers at the intermediate frequency[6], our approach promises quantum–limited sensing in the THz domain, potentially surpassing superconducting technologies, particularly for large heterodyne arrays.**

Observations in the terahertz (or the far-infrared) frequency range (~100GHz-10THz) are of great importance for understanding physics and chemistry in the star and planet forming regions.[7,8] With the wavelength in THz range much longer than that in the IR, terahertz waves have capacity to reveal processes hidden behind dusty gas clouds. Space born telescopes allow overcoming severe atmospheric absorption, whereas a diverse park of instrumentation permits to cover a wide range of scientific tasks. To recover information carried by faint celestial signals, THz frequency mixers -the core of coherent detection- have to fulfill stringent requirements on both sensitivity and, more importantly, on bandwidth, to enable line surveys and studying Doppler-stretched molecular lines. Superconducting hot-electron bolometer (HEB) mixers form the baseline for modern astronomical receivers above 1 THz. In these, the wave beating between the incoming THz signal and detuned local oscillator (LO) causes temperature oscillations at the intermediate frequency (IF), enabling read-out through changes in electrical resistance $R$ (resistive read-out) as long as the temperature in the material can follow the signal modulation. The upper limit for the IF frequency is determined by either the electron-phonon relaxation time $\tau_{e\text{-}ph}$ (phonon cooling) or by the out-diffusion time $\tau_D = L^2/(\pi^2 D)$ of hot electrons from the superconductor into the cold electrical contacts (diffusion cooling), with $L$ the length of the device and $D$ the diffusion constant[9–11]. Despite great efforts, NbN superconducting HEB mixers are limited in terms of instantaneous bandwidth to ~ 4-5GHz[3]. Higher bandwidths are possible in new superconductors[12], but at the expense of forbiddingly high LO power requirements, which is particularly detrimental for array applications. New materials and concepts are needed to go beyond few-pixel THz detectors, available today, to the large-format detector arrays required to enable further cosmic quests[4,5,7].

In this paper we explore the resistive read-out of the bolometric response of graphene, where quantum effects introduce the long-sought-after temperature dependence of graphene's resistance[13–15]. Both long phase coherence times and poor screening of carriers in two-dimensions favor quantum interference and electron-electron interactions[16–18]. These effects introduce logarithmic-in-temperature $T$ dependence of the conductivity $\sigma$ of graphene, of the order of $\sigma_1 \sim e^2/h \approx 3.9 \times 10^{-5}$ S:

$$\sigma(\mathrm{T}) = \sigma_0 + \sigma_1 \ln(T/1K) \qquad (1)$$

With graphene at high carrier density $n$, these effects are relatively small. Yet, as $n$ approaches the Dirac point, its conductance $\sigma_0$ decreases, and the relative contribution of quantum effects grows. This scenario, appealing for the use of graphene as a thermometer, can however be spoiled by charge disorder. Close to charge-neutrality, disorder leads to doping inhomogeneity, resulting in charge puddles which can effectively shunt quantum transport and cause the resistance of graphene to saturate at low temperatures[19–21]. In contrast, for high quality graphene such as flakes encapsulated by boron nitride[22,23], reduced disorder results in monotonically increasing resistance at low temperatures. As we recently demonstrated[24], a similar physical situation occurs when epitaxial graphene on SiC is brought close to the Dirac point by chemical doping. The spontaneous accumulation of assembly of molecular dopants 2,3,5,6-tetrafluoro-tetracyano-quino-dimethane (F4TCNQ) on graphene via diffusion through poly (methyl-methacrylate) (PMMA) results in a molecular adlayer that dopes epitaxial graphene to the vicinity of the Dirac point and flattens charge inhomogeneity, translating into an effectively reduced disorder. As a result, eq. (1) holds even when graphene is doped to the Dirac point ($n < 5 \times 10^9$ cm$^{-2}$) (Fig. 1a), arising primarily from quantum interference in the presence of strong intervalley scattering (Fig. 1b) [16,22], thus making this material an effective thermometer down to $T = 0.2$ K.

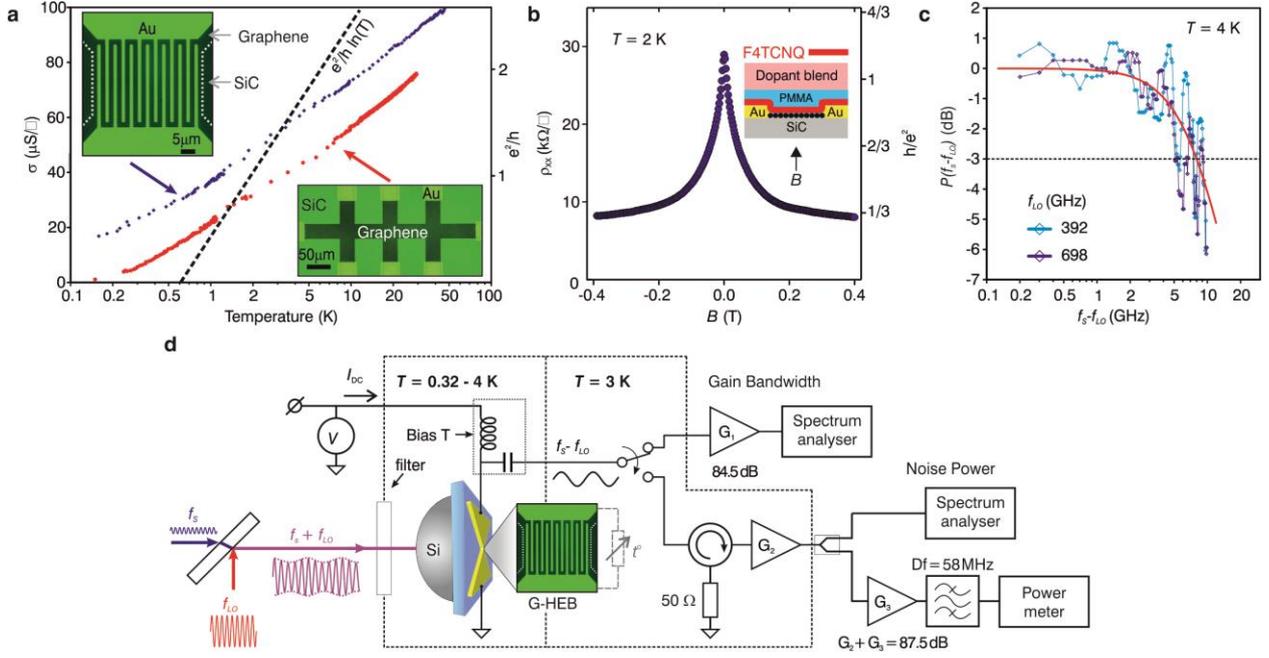

**Fig. 1.** Graphene doped to the Dirac point as bolometric mixer. **b,** Zero bias differential conductivity $\sigma(T) = dI/dV|_{V=0}$, measured in a dark (optically tight) cryostat for a Hall bar (red) and the THz mixer (blue), demonstrating logarithmic temperature dependence $\sigma(T) = \sigma_0 + \sigma_1 \ln(T/1K)$. The dashed line indicates slope $\sigma_1 = e^2/h$. **b,** Suppression of quantum interference effects by perpendicular magnetic field *B* breaking time reversal symmetry (inset shows the cross-section of the device). **c,** The gain bandwidth of 8 GHz for our graphene mixer, extracted via measurements of the power at the intermediate frequency $P_{IF}$ as a function of detuning frequency between the signal $f_s$ and the local oscillator $f_{LO}$. The solid line is a fit to $P(f_{IF}) = 1/[1 + (f_{IF}/f_0)^2]$, the dashed line indicates $P(f_{IF}) = -3$ dB. Resonances are due to interferences in the optical path and the IF chain. **d,** Schematics of the experimental setup: the terahertz signal at frequency $f_s$ is combined with a monochromatic wave emitted by a local oscillator at a nearby frequency $f_{LO}$, and both are fed into graphene via an integrated bow-tie antenna through a silicon lens.

But, from the point of view of bolometric mixer operation, how fast and sensitive can this thermometer be, while based on charge-neutral graphene? Dependence (1) implies a diverging sensitivity of the resistive read-out $dR/dT \sim T^{-1} ln^{-2}(T)$, favoring measurements at low temperatures. The low heat capacity of graphene translates into fast device operation[25] as long as a proper heat-link is provided. However, the electron-phonon cooling time $\tau_{e-ph} \sim n^{-0.5} T^{-2}$, diverges at charge neutrality and low temperature[26,27]. Therefore, other cooling pathways, such as electron diffusion cooling, will be required. Although the residual charge puddles do not shunt quantum interference effects, the exact chemical potential landscape and microscopic scattering details are not known for the studied material, and these may greatly modify charge and heat transport of graphene at the Dirac point [28–31]. Summarizing, it is a priori not clear how quantum interference and strong electron-electron interactions in epitaxial graphene, chemically doped to the vicinity of the Dirac point, will affect the device speed and noise - it has to be tested experimentally.

We conducted THz mixing experiments to unveil both the response time and mixer sensitivity in epitaxial graphene chemically doped to the Dirac point. Figure 1c, the central result of this experiment, shows a mixing gain bandwidth of $f_0 = 8$ GHz, corresponding to a time constant of $1/(2\pi f_0) = 20$ ps (for temperature independence of gain bandwidth see fig. S1). As we elaborate below, the high Fermi velocity in graphene and highly transparent metallic contacts[32] in our device allow the heat load in the device to be quickly dissipated by hot electrons diffusing into the metallic leads. To match the resistance of the graphene sample to both the THz impedance of the antenna and the impedance of the IF readout amplifier, we used interdigitated electrodes, with a graphene length of $L = 1.5$ um and a device width of $W = 345$ um, resulting in nominally 230 squares of graphene in parallel. On-chip Hall bar devices ($L = 180 \ \mu m \ \text{x} \ W = 30 \ \mu m$) display similar conductivity and temperature dependence $\sigma(\text{T}) = \sigma_0 + \sigma_1 \ln(T/1K)$, confirming our device retains the properties of the 'bulk' chemically doped graphene (Fig. 1a, fig. S2). Figure 1d shows a schematic of our setup, which simultaneously allows for direct current (DC), THz response, and noise thermometry measurements down to $T = 0.32$ K (see also fig. S3). Thorough calibration of the THz power reaching graphene and the gain of the GHz chain was performed using noise thermometry (fig. S4, S5).

Remarkably, the DC transport characteristics, and both the THz direct and mixing response of our device –with transport properties dominated by quantum effects- can all be understood within a simple thermal model of diffusion cooling of electrons. In essence, the effect of THz radiation on our device is that of a temperature increase, with a diffusion cooling rate dominating over phonon cooling of hot electrons (see Supplementary Section 1 and fig. S6). Figure 2a shows the differential resistance (*dV/dI*) of the device, measured in the optical cryostat with the THz source OFF, as a function of the bias current at different temperatures. Complementary, Fig.2b shows *dV/dI* of the device at the base temperature $T_0 = 0.32$ K for increasing THz power. In both figures, solid lines correspond to calculations within a Diffusion Cooling Model (DCM), which considers that the heat induced by both DC current and THz radiation is transferred by charge carriers into the metallic leads kept at the cryostat temperature $T_0$ (see also fig. S6, S7 and Supplementary Section 2). Our model assumes the validity of Wiedeman-Franz law and considers that, for temperatures below 6 K, phonon cooling of electrons can be neglected. For the particular case of logarithmic temperature dependence of the bridge conductance, the device current-voltage characteristic *I(V)* under a THz power $P_{ac}$ can be obtained within the DCM as a recursive expression:

$$I(V) = G(T_0)V + G_1 V \left( \frac{\sqrt{1+u^2}}{u} \ln\left(\sqrt{1+u^2} + u\right) - 1 \right) \quad (2)$$

with $u = \frac{1}{V_T}\sqrt{V^2 + \frac{V}{I}P_{ac}}$ and $V_T = \sqrt{\mathcal{L}} \cdot T_0$. Here, $G(T) = G_0 + G_1 \ln(T/1K)$ is the experimentally determined temperature-dependent conductance, yielding $G_0 = 6.55$ mS and $G_1 = 3.50$ mS. From the family of differential resistance curves measured with the THz source OFF in Fig. 2a, we extract as a fitting parameter the radiation background power $P_{Bkg} = 0.28$ nW in our cryostat, which leads to apparent saturation of conductance at low temperatures (Fig. 2a, inset) . Additionally, the fit returns the Lorenz number $\mathcal{L} = 3.1 \times 10^{-8}$ WΩ/K² remarkably close to its theoretical value, $\mathcal{L} = \pi^2/3\, (k_B/e)^2 \approx 2.44 \times 10^{-8}$ WΩ/K², despite graphene being at the charge neutrality point[33]. Next, we use the same $P_{Bkg}$ and $\mathcal{L}$ to predict a set of curves for the *dV/dI* response under THz irradiation at base temperature $T_0 = 0.32$ K, considering that the total absorbed THz power ($P_{ac}$) includes contribution from both the LO and $P_{Bkg}$. Figure 2b shows, with no extra fitting parameters involved, that the DCM precisely reproduces the observed bolometric response to radiation.

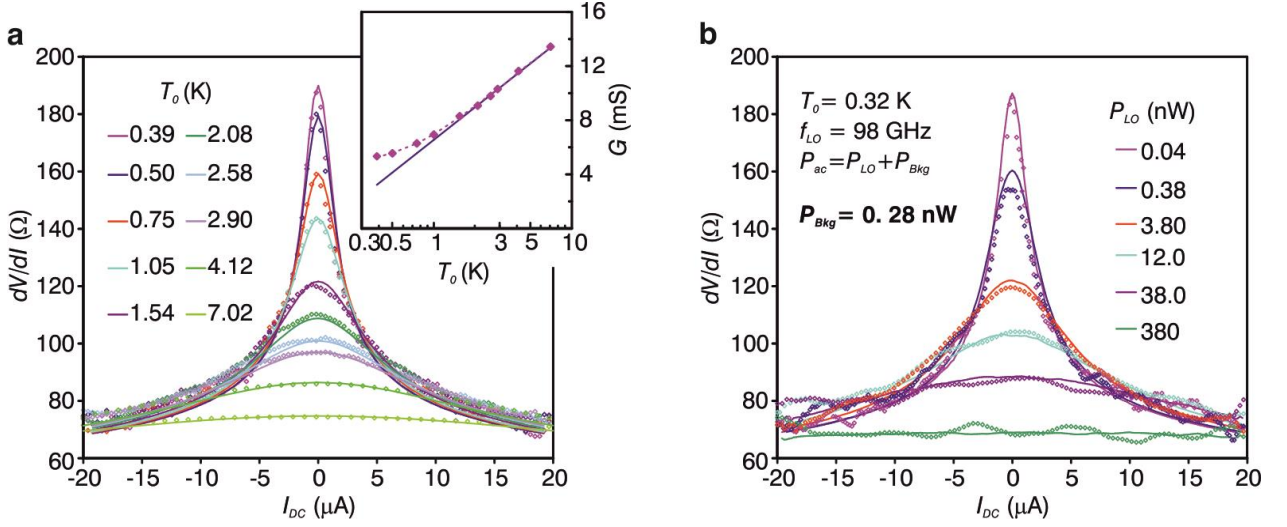

**Fig. 2.** Equivalence of THz radiation, Joule heating and base temperature effects on the graphene bolometer. **a,** Effect of temperature on the measured differential resistance $dV/dI(I)$ with the THz source OFF. Inset: Zero bias conductance for the same device. Deviation from $G(T) = G_0 + G_1 \ln(T/1K)$ at low temperature, absent in measurements in a dark cryostat (Fig. 1A), is due to background heat load $P_{Bkg}$. **b,** Effect of THz power ($f_{LO} = 98$ GHz) on the $dV/dI$ response of our device at $T_0 = 0.32$ K. In both cases dots correspond to measurements and solid lines to the DCM prediction.

Having established the bolometric nature of the THz response in our device, we investigate its performance as a THz mixer. We irradiated the device with two THz sources, a signal S, and the LO, measuring the mixer gain $G_{mix} = P_{IF}/P_S$ which is the ratio of the power at the output ($P_{IF}$) to the incoming signal power $P_s$ (Fig. 3A). Solid lines in Fig. 3a show that, within all of our tested parameter space ($T_0, P_{LO}, I_{DC}$), $G_{mix}$ is well described by the classical bolometric mixer expression [34], which also follows from the DCM (for details of the DCM, see Supplementary Section 2):

$$G_{mix} = \frac{1}{2}\frac{P_{LO}}{P_{DC}}\frac{50\Omega}{V/I}\left[\frac{1-\left(\frac{V}{I}\right)\left(\frac{dI}{dV}\right)}{1+50\Omega\left(\frac{dI}{dV}\right)}\right]^2 \qquad (3)$$

The maximum mixer gain is $G_{mix} = -27$ dB ($P_{IF}/P_S = 2\%$) at optimum mixing conditions ($P_{LO} = 3.8$ nW, $I_{DC} = 5$ μA). At this operating point, the electronic temperature in graphene, measured by noise thermometry, is elevated above the base temperature due to the DC bias, the LO, and background radiation powers, reaching an effective noise equivalent temperature $T_S \approx 1.9$ K, as shown in Fig. 3b. Solid lines in Fig. 3b correspond to the DCM considering that the microwave noise of graphene is purely thermal (Johnson noise) (see also Supplementary Section 3 and fig. S8). For a coherent mixer, the ultimate sensitivity is determined by its noise temperature $T_{mix}$, and for our device we find $T_{mix} = T_S/(2G_{mix}) = 475$ K (a factor 2 accounts for the double sideband nature of coherent mixing).

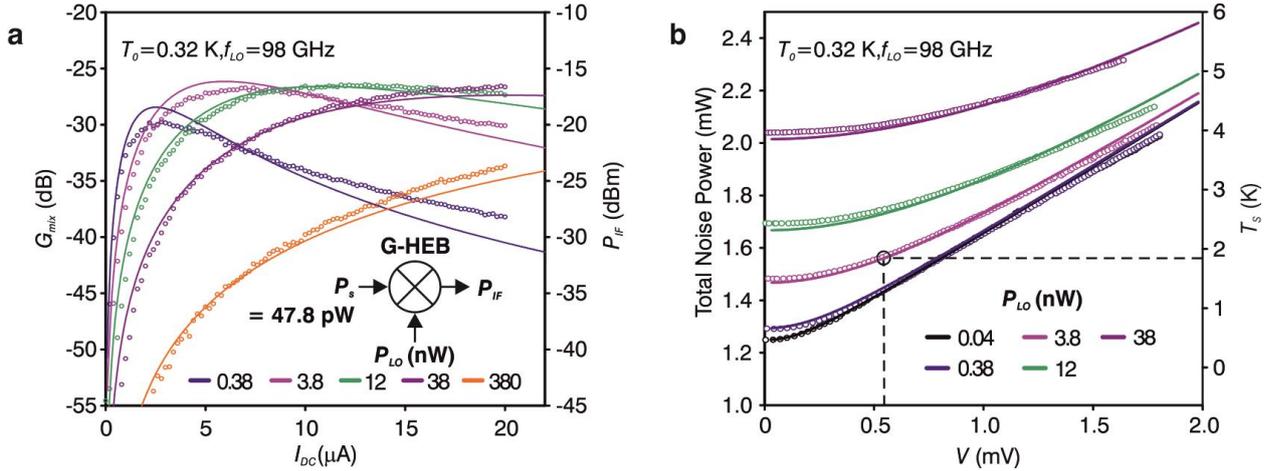

**Fig. 3**. Terahertz mixing performance of our graphene bolometer. **a,** Mixing gain $G_{mix}$ ($f_{IF} = 3.5$ GHz, $P_S = 47.8$ pW) and **b,** Thermal noise power measurements for different LO power levels $P_{LO}$ ($f_{IF} = 4$ GHz). Dots correspond to measurements. With a radiation background $P_{Bkg} = 0.5$ nW, solid lines in **(a)** correspond to eq. (2) and the right y-axis is the measured output power at the IF. In **b**, solid lines are fits to the DCM, and the right y-axis displays the graphene noise equivalent temperature $T_S$, calculated with the DCM.

With the validity of DCM demonstrated, we extrapolate the intrinsic performance of our device for zero background radiation conditions ($P_{Bkg} = 0$), as expected on a space-born mission. Figure 4a shows that zero background allows for operation at reduced DC bias, resulting in the increase of the achievable mixing gain $G_{mix}(I_{DC}, P_{LO})$ by a few dB compared to the gain we measured in our cryostat (Fig. 3A). Combining the mixer gain and the $T_S$ corresponding to maximum conversion conditions, we show the calculated mixer noise temperature as a function of the base temperature $T_0$ in Fig. 4b (see also Supplementary Section 4 and fig. S9). Under these considerations, cooling the device down to $T_0 = 0.2$ K, a limit set by the validity of logarithmic fit in our model and achievable in refrigerators for satellite missions[35], leads to $T_{mix} = 36$ K. Remarkably, this implies that the graphene bolometer mixer will operate in a quantum-limited regime at all frequencies above $k_B T_{mix}/h = 0.75$ THz.

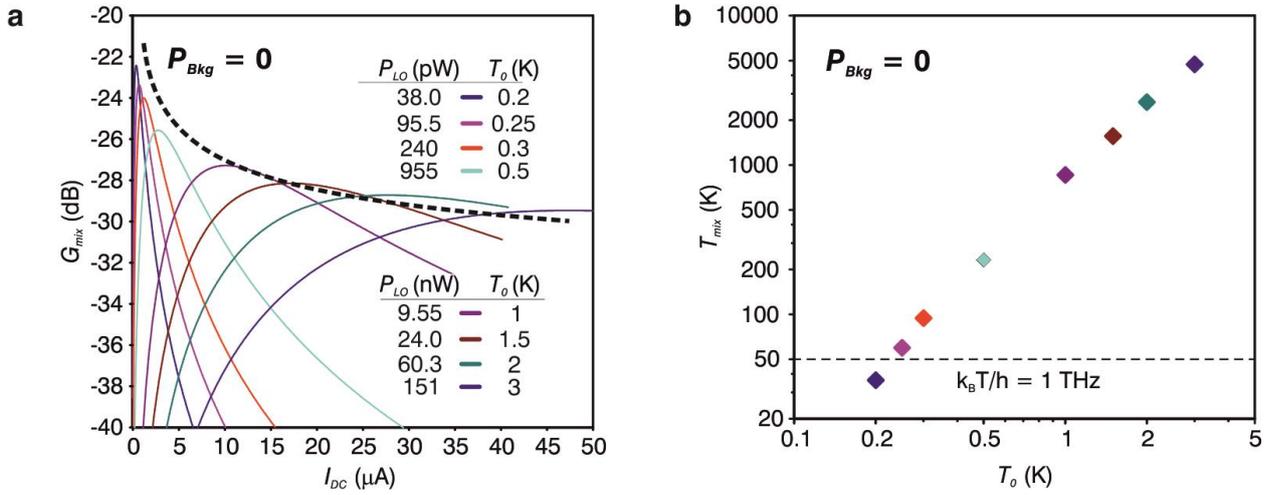

**Fig. 4**. Intrinsic performance of the graphene bolometric mixer (without background radiation). **a,** Mixing performance for different operating temperatures $T_0$. For each temperature the curve corresponding to an optimum LO power (highest possible gain) is shown. Dashed line represents an approximated expression for the mixing gain $G_{mix} \propto (G_1 V/I)^2$. **b,** Mixer noise temperature ($T_{mix} = T_{noise}/2G_{mix}$) under optimal conditions as a function of the base temperature $T_0$.

The performance of our THz mixer improves drastically at lower temperatures, and thus can take full advantage of space-oriented cryocoolers providing sub-Kelvin temperatures[35]. In our experiments the graphene temperature at optimum operation conditions was limited by the background radiation power. At zero radiation background, we expect $T_S \sim 0.4$ K; combined with the best available commercial amplifiers ($T_{Amp} \sim 1$ K), this translates into a mixing noise figure of 125 K for the mixer-amplifier chain. Unleashing the full potential of the presented mixer requires a wideband amplifier with sub-Kelvin noise. Such devices are emerging on the wave of progress in quantum technologies[6]. For an IF amplifier with an added noise of about one photon at 10 GHz, the mixer input-referred noise will be $1/G_{mix} \approx 100$ photons, or $\sim 1$ in units of 1 THz photons.

Our results are based on our ability to decrease the carrier density in graphene without degrading its transport properties. In this scenario, quantum transport in graphene at the Dirac point enables highly sensitive and broadband terahertz coherent mixing. Currently demonstrating a gain bandwidth of 8 GHz for a $L = 1.5$ μm long device, and with its operation fully described by a diffusion cooling model, we expect the bandwidth of the device to scale as $\sim L^{-2}$ (see also Supplementary Section 1 and fig. S6)[9–11]. Thus, for a device length of $0.8 - 1$ μm (attainable with current microfabrication technologies), the bandwidth can in principle be extended to 20 GHz, far superior to what is achievable in superconducting mixers. Combining wafer-scale epitaxial graphene on SiC and sub-nW LO power requirements, this mixing platform is appealing for implementing large focal plane arrays of THz mixers. Finally, the achievable mixing gain $G_{mix} \propto (G_1 V/I)^2$ is set by the strength of quantum effects, which in our material are quantified as $\sigma_1 \approx 0.3 \ e^2/h$ (Fig. 1a). Further efforts will unveil if a different combination of polymers and molecular dopants can lead to an increased $\sigma_1$, as a result of an even more homogeneous doping closer to the Dirac point or by tuning a particular interplay of microscopic scattering mechanisms.

refrigerator insensitive to gravity. in *Cryogenics* **50,** 623–627 (Elsevier, 2010).

**Acknowledgments**: We thank Alexander Tzalenchuk, Justin F. Schneiderman and Tord Claeson for useful discussion and critical reading of the manuscript. This work was jointly supported by the Swedish Foundation for Strategic Research (SSF) (No. IS14-0053, GMT14-0077, RMA15-0024), Knut and Alice Wallenberg Foundation, Chalmers Area of Advance NANO, the Swedish Research Council (VR) 2015-03758 and 2016-048287, the Swedish-Korean Basic Research Cooperative Program of the NRF (No. NRF-2017R1A2A1A18070721), European Union's Horizon 2020 research and innovation programme (grant agreement No 766714/HiTIMe).

**Author contributions:** S.L.A., H.H., K.H.K. and R.Y contributed to sample growth and device fabrication. S.L.A., H.H., K.H.K, F.L. and T.B. performed the DC characterization of the device. D.G. developed the theoretical calculations. A.D. and S.Ch. characterized the sample at THz and microwave frequency ranges. D.G., S.L.A, A.D., S.Ch and S.K contributed to the interpretation of the experiments. S.K., S.L.A and S.Ch. conceived and designed the experiment. All the authors contributed to the writing of the manuscript.

**Competing interests:** Authors declare no competing interests.

**Materials & Correspondence:** should be addressed to samuel.lara@chalmers.se

**Data and materials availability:** The authors declare that the main data supporting the findings of this study are available within the article and its Supplementary Information files. Additional data is available from the corresponding author upon request.

**Methods**

<u>Sample Fabrication</u>

Monolayer SiC/G was grown on the Si-face of SiC using thermal decomposition of 7 mm x 7 mm SiC substrates. The samples were grown in argon atmosphere 800 mbar, and at a temperature of around 1700 ˚C for 5 min. Optical microscopy reveal a typical surface coverage of > 95 % monolayer graphene. Devices were fabricated using conventional electron beam lithography, using only PMMA resist (Microlithography Chemicals Corp.) bilayer and lift-off. The electrical contacts were fabricated using physical vapor deposition of Ti/Au, 5 nm and 80 nm thick respectively. After fabrication, samples were cleaned using isopropyl alcohol, acetone and dried using nitrogen gas. After this step, the samples were spin-coated with a 100 nm-thick PMMA layer (Microlithography Chemicals Corp.), followed by spin coating of a 170 nm-thick chemical dopant blend. The dopant blend consists of a mixture of F4TCNQ (Sigma-Aldrich) and PMMA (Microlithography Chemicals Corp.). In detail, 25 mg of dry F4TCNQ powder is mixed with 3 ml anisole solvent. Subsequently, 0.5 ml of this solution is then mixed with 1 ml PMMA A6 (6% PMMA by weight in anisole). The resulting ratio of PMMA to FT4CNQ is roughly 93:7 by weight. All polymer layers are deposited on graphene using spin coating at 6000 rpm for 1 min. A 5 min baking step on a hotplate at 160 ˚C follows each spin coating step. This process consistently yields a carrier density $|n| < 1 \times 10^{10}$ cm$^{-2}$ and sheet resistance of about $\rho = 25$ k$\Omega$/sq at $T = 2$ K.

<u>Gain Bandwidth measurements</u>

Frequency mixing experiments were done with the two-wave mixing approach, where the frequency of one THz source is kept constant, while the frequency of the second source is changed. The resulting THz current through the device becomes amplitude modulated with the beating frequency ($f_{IF} = |f_{LO} - f_S|$). The output signal from the device under test is amplified and its power is measured with a spectrum analyzer. Measurements were conducted with THz generators in all three frequency bands 100, 300 and 700 GHz. For broadband measurements, with the beating frequency $f_{IF}$ at least up to 10 GHz, both the bias-T and the amplifiers were held at room temperature. Gain calibration for the readout chain was conducted separately. The only factors which were not included in the calibration were the effect of the mixer block and the THz standing waves in the optical path. These residuals lead to some ripples in the measured signal. The fitting curve $P(f_{IF}) = 1/[1 + (f_{IF}/f_0)^2]$ in Fig. 1C of the main text provides a roll-off frequency of $f_0 = 8$ GHz.